# ENUM: The Collision of Telephony and DNS Policy

By Robert Cannon
cannon@cybertelecom.org

***Disclaimer:*** *Views expressed are probably those of Robert Cannon and certainly are not necessarily those of anyone else, including, but not limited to, Robert Cannon's employer.*

## Table of Contents

# Introduction

ENUM marks either the convergence or collision of the public telephone network with the Internet. ENUM is an innovation in the domain name system (DNS). It starts with numerical domain names that are used to query DNS name servers. The servers respond with address information found in DNS records. This can be telephone numbers, email addresses, fax numbers, SIP addresses, or other information. The concept is to use a single number in order to obtain a plethora of contact information.

By convention, the Internet Engineering Task Force (IETF) ENUM Working Group determined that an ENUM number would be the same numerical string as a telephone number. In addition, the assignee of an ENUM number would be the assignee of that telephone number. But ENUM could work with any numerical string or, in fact, any domain name. The IETF is already working on using E.212 numbers with ENUM.

ENUM creates multiple policy problems. What impact does ENUM have upon the public telephone network and the telephone numbering resource? For example, does this create a solution or a problem for number portability? If ENUM truly is a DNS innovation, how does it square with the classic difficulties experienced with DNS and ICANN? Is ENUM, while presenting a convergence solution, also encumbered with the policy problems of both the DNS and telephony worlds?

IETF ENUM proponents suggest that ENUM needs a single unified database administered through national and international government sanctioned monopolies. The IETF took the unusual step of requesting that the International Telecommunications Union (ITU) regulate an aspect of the Internet, that is, participate and have authority over the international ENUM service provider. But this notion of establishing a new communications monopoly collides with the deregulatory efforts of the Telecommunications Act of 1996, the attempts to privatize DNS through ICANN, and US policy that the Internet should be left unregulated. ENUM is an unproven innovation with no evidence of commercial viability. It faces a strongly competitive market of other directory assistance innovations and services. Proponents are asking governments to sanction one competitor over others.

ENUM offers two lessons. First, involving the government in a standards process is fraught with problems and delays. It starts with the cliché of having too many cooks in the kitchen, producing a mediocre cake at best. And it ends with a cumbersome bureaucratic process resulting in fatal delay and ultimately collapsing in upon itself. Similar efforts in the past rose to grandiose levels and failed. These include X.500 and OSI.

Second, a number by any other name remains a number. A significant portion of the DNS wars has been focused on resolving who has the right to a name. Is it first come, first serve, a trademark holder, someone using the domain name pursuant to free speech rights, or perhaps some other right? With ENUM, the question presented is who has the right to a numerical string. ENUM attempts to resolve this question by convention, concluding that the assignee of a telephone number has rights to an ENUM number. But an ENUM number is not a telephone number. A telephone number is an address used on a telephone network to reach a telephone. An ENUM number is a token used to access a database. Transferring a numerical string from one context to another does not likewise transfer the rules and regulations of the original context. Rules and regulations created for telephone numbers assume a particular purpose in a particular context; they do not apply to numerical strings in a foreign context with a different purpose. It is illogical and dangerous to transfer the policy concerning one type of number to a different type of number. This means, among other things, that the regulatory authority over telephone numbers has no more jurisdiction over ENUM numbers then when telephone numbers are used to rent videos or access savings clubs at the grocery store.

US policy has been to keep information technology unregulated to permit it to innovate at the speed of the market and not at the pace of bureaucracy. Yet ENUM proponents beg for government entanglement. It would be unprecedented for the government to sanction a monopoly for something as unproven as ENUM where the appropriateness of a government monopoly has not been demonstrated. Were such government involvement in fact approved, the delay experienced would likely be fatal to the innovation.

There are those who are strong advocates of an ENUM unified database. An ENUM unified database can likely be achieved by private industry through some level of a joint venture devoid of government entanglement. This is the best hope for ENUM achieving the goal of a swift implementation.

## ENUM

ENUM is an IETF proposed standard[1] (RFC 2916[2]) created by the IETF ENUM Working Group.[3] It is an Internet domain name system (DNS) innovation.[4]

---

[1] *See* S. Bradner, IETF RFC 2026, The Internet Standards Process -- Revision 3 (October 1996) (hereinafter RFC 2026) (explaining IETF process and difference between proposed, draft, and Internet standards), *at* http://www.ietf.org/rfc/rfc2026.txt.

[2] P. Faltstrom, IETF RFC 2916, E.164 number and DNS (September 2000) (hereinafter RFC 2916), *at* http://www.ietf.org/rfc/rfc2916.txt. *See also* Report of the Department of State ITAC-T Advisory Committee Study Group A Ad Hoc on ENUM (Jul. 6, 2001) (hereinafter Ad Hoc ENUM Report) (presenting US industry views to US State Department concerning implementation of ENUM), *at* http://www.cybertelecom.org/library/enumreport.htm.

[3] *See* IETF ENUM Working Group Charter (last visited August 14, 2001), *at* http://www.ietf.org/html.charters/enum-charter.html.

[4] *See* Ad Hoc ENUM Report, *supra* note 2, Sec. 2 (stating ENUM is a protocol whereby "'Domain Name System (DNS) can be used for identifying available services connected to one E.164 number.'"), *at* http://www.cybertelecom.org/library/enumreport.htm; Contribution of NeuStar, Inc.,

Personal contact information within DNS records can be retrieved using an ENUM number. A ENUM number is entered, it queries the a DNS name server which then responds with telephone numbers, IP telephony numbers, fax numbers, e-mail addresses, and telephone number after 5:00 p.m. on weekends.[5] It can also provide information about the priority pursuant to which the record owner wishes to be contacted. Thus, having only a single identifier, a user could acquire all of the contact information for an individual.[6]

ENUM numbers are converted by ENUM devices into domain names, and then used to query the domain name system. If an ENUM record exists, then the database produces the contact information. The ENUM device is on the Internet, the query is over the Internet, and the ENUM database is on the Internet. It can be used in conjunction with a multitude of applications on or off the Internet including telephony, email, fax, and others.[7]

---

US Study Group A Ad-Hoc, ENUM Questions, p. 5 (March 23, 2001) (hereinafter NeuStar, Inc., US Study Group A Ad-Hoc,) (stating "ENUM is a DNS-based service"); NeuStar, ENUM Frequently Asked Questions, FAQ-7 (n.d.) (hereinafter NeuStar FAQ) (stating "This is a DNS-based system…"), *at* http://www.enum.org/information/files/enum_faq.pdf; S. Lind, IETF Informational Internet Draft, ENUM Call Flows for VoIP Interworking, para 2 (Nov. 2000) (hereinafter Lind, Callflows) (stating "ENUM provides the capability to translate an E.164 Telephone Number into an IP address or URI using the Domain Name System (DNS)"), *at* http://www.ietf.org/internet-drafts/draft-lind-enum-callflows-01.txt; Penn Pfautz, James Yu, IETF Informational Draft, ENUM Administrative Process, Sec. 1 (March 2001) (hereinafter Pfautz, ENUM Administrative Process) (stating "after all it is a domain name that is being registered"), *at* http://www.ietf.org/drafts/draft-pfautz-yu-enum-adm-01.txt. *See also* Richard Shockey, IETF-ENUM ITU-T Workshop for International Regulators, slide 7 (January 17, 2001) (hereinafter Shockey, ITU-T) (explaining reason for placing ENUM in DNS is "It's there… It works… It's global… It scales… It's fast… It's open."); A. Brown, G. Vaudreuil, IETF Internet Draft, ENUM Service Reference Model, Sec. 5.1 (Feb. 23, 2001) (hereinafter, Brown, ENUM Service Reference Model) (stating "The Internet Domain Name System provides an ideal technology for the first-tier directory due to its hierarchical structure, fast connectionless queries, and distributed administrative model."), *at* http://www.ietf.org/internet-drafts/draft-ietf-enum-operation-02.txt.

This article relies primarily on primary sources in the ENUM policy debate. These sources are on file with the author. Most Internet documents are linked at http//:www.cybertelecom.org/enum.htm.

[5] Ad Hoc ENUM Report, *supra* note 2, Sec. 6.1, *at* http://www.cybertelecom.org/library/enumreport.htm.

[6] *See* Lind, Callflows, *supra* note 4, para 2 (noting ability to change contact information without changing ENUM number).

[7] In addition, it has been discussed that instead of having addressing information in the NATPR record, the NAPTR would point to a third-party database such as the LDAP database. Such a NAPTR record could look like "IN NAPTR 10 10 "u" "Reachme+E2U" \ "!LDAP:\\dap1.zcorporation.com\cn=\!" . *See* Ad Hoc ENUM Report, *supra* note 2, Sec. 5.2.2, *at* http://www.cybertelecom.org/library/enumreport.htm; Brown, ENUM Service Reference Model, supra note 4, Sec. 7.1, *at* http://www.ietf.org/internet-drafts/draft-ietf-enum-operation-02.txt.

```
Sample ENUM DNS Record:

$ ORIGIN 2.1.2.1.5.5.5.2.0.2.1.1.E164.foo[8]
    IN NAPTR 102 10 "u" "tel+E2U" "!^.*$!tel:+112025551212!"  .
    IN NAPTR  10 10 "u" "sip+E2U" "!+(.*)!sip:johndoe@company.com!"  .
    IN NAPTR 100 10 "u" "mailto+E2U" "!^$!mailto:johndoe@company.com!"  .
```

The IETF ENUM WG determined that ENUM numbers would have the same value as a person's telephone number. The assignee of a telephone number would be the assignee of an ENUM number.[9] This achieves several goals. It creates a global standard form for ENUM numbers - they could be anything. It creates a standard for how ENUM numbers shall be assigned. It also means that ENUM numbers, which are domain names, are numeric (unlike most domain names which utilize letters and words), can be entered into telephone number pads, are linguistically neutral, and can take advantage of the familiarity of the public with telephone numbers.[10]

ENUM would function as follows: A user in Washington, D.C. may wish to reach the reach the Joe.
- The user inputs into an ENUM enabled device the ENUM number 555-1212.
- The ENUM device expands the ENUM number into the same numerical string as the full E.164 number: 1-1-202-555-1212.[11]

---

[8] The use of "foo" as a TLD is an informal IETF convention indicating that the TLD is unspecified. *See* D. Eastlake, C. Manros, E. Rayond, IETF Information RFC 3092, Etymology of "Foo" (April 1, 2001) (explaining origins and use of term "foo" in IETF documents; "foo" is used "as a sample name for absolutely anything, esp. programs and files."), *at* http://www.ietf.org/rfc/rfc3092.txt.

[9] Several presentations describe the purpose of ENUM as being a means of finding a device on the Internet using a telephone number. *See* Shockey, ITU-T, *supra* note 4, Slide 5; ENUM.ORG > Welcome Page (visited March 27, 2001) ("ENUM was developed as a solution to the question of how to find services on the Internet using only a telephone number, and how telephones, which have an input mechanism limited to twelve keys on a keypad, can be used to access Internet services.") *at* http://www.enum.org; Patrik Faltstrom, ENUM Technical Issues, ITU ENUM Work Shop, slide 12 (Jan 17, 2001) (hereinafter Faltstrom, ENUM Technical Issues). However, the ENUM database can contacted personal and contact information for all types of devices and locations, not just Internet devices.

[10] *See* NeuStar FAQs, *supra* 4, FAQ-1 (stating ENUM was designed to permit access to Internet services using a telephone keypad), *at* http://www.enum.org/information/files/enum_faq.pdf; Richard Shockey, IETF-ENUM SGA-Workshop on ENUM, slide 9 (n.d.) (hereinafter Shockey, SGA).

[11] E.164 is the international telephone numbering plan administered by the ITU. *See* Recommendation E.164/I.331 (05/97) - The International Public Telecommunications Numbering Plan, *at* http://www.itu.int/itudoc/itu-t/rec/e/e164.html; Robert Shaw, ITU, Global ENUM Implementation, DTI ENUM Workshop, Slide 3 (June 5, 2001) (hereinafter Shaw, DTI ENUM Workshop), *at* http://www.itu.int/infocom/enum/dtijune501/dti-june-5-2001-1.PPT; Robert Shaw, ITU, ENUM Implementation, ICANN Governmental Advisory Committee, Slide 3 (1-2 June 2001) (hereinafter Shaw, ICANN), *at* http://www.itu.int/infocom/enum/GACjune1201/gac-june-2-2001-1.PPT.

- The ENUM device reverses the number, removes non-number symbols, and converts the number into a domain name. The device would create the ENUM number domain name <2.1.2.1.5.5.5.2.0.2.1.1.foo>.
- This domain name would then be sent to a designated ENUM name server on the Internet. A DNS query would be conducted for each zone of the domain name.[12]
- If a record exists, the database would produce the result that could, for example, direct the user first to call Joe's IP telephony number, second to contact Joe's e-mail address, or finally to call Joe's number.[13] The result would also reflect the preference of the person on how that person prefers to be contacted.[14] If no record exists, the user will receive an error message similar to receiving an error message when requesting a webpage that does not exist.[15]
- Based on the user, the person the user desires to contact, and the ENUM information provided, the communication would then be set up by other applications (not by ENUM). If the information used is a URI, an additional DNS lookup must be conducted to get the IP address.

In order for ENUM to work, there must be an ENUM enabled device. All a device would need is a bit of software, meaning any device capable of running the software that has Internet access could be enabled. The device would receive the ENUM number, convert it into a domain name, and then conduct the query. It is edge technology at either the originator's or the terminator's edge. ENUM devices would be programmed to point to a designated Internet name server[16]

---

[12] *See* Brown, ENUM Service Reference Model, *supra* note 4, Sec. 6.1, at http://www.ietf.org/internet-drafts/draft-ietf-enum-operation-02.txt.

[13] *See* NeuStar FAQ, *supra* note 4, FAQ-1 (stating that "once the authoritative name server is found, ENUM retrieves relevant NAPTR Resource records . . ."), *at* http://www.enum.org/information/files/enum_faq.pdf. NAPTR stands for "Naming Authority Pointer."

[14] *See Id.*, FAQ-1 (stating user can specify preferences for receiving communications).

[15] *See Id.,* FAQ-5.

[16] Generally, when acquiring Internet services, a user acquires software from the ISP with preconfigured software. The software generally has a series of fields for such data as the users name, e-mail address, and other values. There is generally two fields for name servers, where the IP number of two different servers is stored. When the user sends data using a domain name, the software consults the pre assigned name server for the IP number associated with that domain name. If the first name server fails, the second name server can be consulted. If the second name server fails, the address cannot be resolved and the communication fails. Generally, while the name server fields are configured by the ISP, they can be reconfigured by the user. The user can point to and receive data from any name server the user chooses. Alternative DNS systems have been developed and, in order to participate, all users had to do was enter the IP number of the alternative DNS system and point to the alternative name server. In this manner, unofficial domains such as *.web* can work.

where it would have access to an ENUM database. Either the vendor or the user could program the device.[17]

### *ENUM Administration*

One of the central ENUM issues is how will the database be administered. This issue marks an area of significant contention within the ENUM community. There is strong consensus in favor of the technical aspects of the protocol, however, consensus with regard to ENUM administration does not appear to exist.

Pursuant to RFC 2916[18] and the ITU ENUM Liaison,[19] the database is to be administered in a hierarchical model with a single international database pointing to single national databases for each telephone country code, that in turn point to authorized service providers. This model is broken down into tiers, with Tier 0 being the international level, tier 1 being the national level, and tier 2 being the competitive service provider levels. The hierarchical model is being actively discussed by the ENUM industry and the ITU, and is evolving.

Tier 0: The administrative contact for the international database is the Internet Architecture Board[20] and the technical contact is RIPE NCC.[21] The international database administered by RIPE NCC will be located in the *E164.arpa* domain.[22] The ITU will supply information on the E.164 database, encourage member states to participate, indicate to RIPE-NCC who the authorized provider of a member state is (recognize the credentials of national service providers), and have a vague level of authority.[23] RIPE NCC, having been informed by the ITU what the E.164 numbers are and who should be recognized at the national level, will populate the database only as instructed and authorized by the nation (lacking authorization from a nation, the database will not be populated[24]). The RIPE-NCC database will point to the national database (a.k.a., Tier 1); it would appear that this is the limit of the scope of RIPE-NCC's role and that its database will not contain additional information.[25] Tier 0 would not know about service-specific information associated with individual ENUM numbers.[26]

---

[17] For a description of potential call flows, *see* Lind, Callflows, *supra* note 4, *at* http://www.ietf.org/internet-drafts/draft-lind-enum-callflows-01.txt; Ad Hoc ENUM Report, *supra* note 2, Sec. 6.2, *at* http://www.cybertelecom.org/library/enumreport.htm..

[18] RFC 2916, *supra* note 2, *at* http://www.ietf.org/rfc/rfc2916.txt.

[19] ITU, Liaison to IETF/ISOC on ENUM (October 2000) (hereinafter Liaison), *at* http://www.itu.int/infocom/enum/wp1-39_rev1.htm. *See also* IETF Informational RFC 3026, Liaison to IETF/ISOC on ENUM (January 2001) (hereinafter RFC 3026), *at* ftp.rfc-editor.org/in-notes/rfc3026.txt. Note that an informational RFC is an informational vehicle only and does not indicate the recommendation or endorsement of the IETF. RFC 2026, *supra* note 1, *at* http://www.ietf.org/rfc/rfc2026.txt.

[20] The IAB is a technical advisory group, under the corporate structure of the Internet Society, that provides leadership for the IETF. The IAB selects the IETF's Internet Engineering Steering Group which in turn selects the leadership of the different IETF working groups. The IAB also provides oversight of the standards process and a forum for appeals concerning the process. *See* Internet Architecture Board Home Page (last modified Dec. 4, 2000), *at* http://www.iab.org/iab/.

[21] E164.ARPA InterNic WHOIS Record (last modified June 22, 2001); E164.ARPA Network Solutions WHOIS Record (last modified Mar. 13, 2001). RIPE NCC is one of three high level Internet numbering authorities. It receives number blocks from the Internet Assigned Number Authority (IANA) which is under the authority of the Internet Corporation for Assigned Names and Numbers (ICANN). It distributes numbers to networks in Europe and Africa. RIPE NCC is located in the Netherlands. About RIPE (n.d.), *at* http://www.ripe.net/ripe/about/index.html.

[22] *See* Shockey, ITU-T, *supra* note 4, slide 11 (explaining that IAB selected *.arpa* because *.arpa* is dedicated to infrastructure issues and is well managed, state and secure).

Tier 1: National ENUM Service Provider are to be set up by a national regulatory authority, possibly through a procurement process.[27] It would be a government sanctioned monopoly, designated to the ITU as the Tier 1 provider.[28] The Tier 1's role is to point to the Tier 2 providers where the actual Naming Authority Pointer (NAPTR[29]) records are retained and authentication of data occurs. Tier 1 does not interact directly with end users.[30]

Tier 2-3: The lower tiers would be comprised of competitive registries who interact with customers and users. They would create, authenticate, and hold the NAPTR records.[31]

End User: Implicitly at the bottom of this model is the end-user. The end-user is the ENUM number assignee and telephone number assignee who is able to create an ENUM DNS record and enter information into the NAPTR records. As the Internet Corporation of Assigned Names and Numbers (ICANN) regulates by contract,[32] requiring all domain name registrants to agree to certain terms, ENUM registrants may be bound by certain terms and conditions of the Tier-1 ENUM service provider including dispute resolution.[33] Registrants could update their records to reflect changes, but if the information is held in the DNS NAPTR records, the information could not be updated in real time. It could only be updated at the speed of DNS refresh.[34]

---

[23] *See* footnote 113, and accompanying text (discussing expanding role of ITU in ENUM).

[24] Liaison, *supra* note 19, *at* http://www.itu.int/infocom/enum/wp1-39_rev1.htm. The Liaison indicates that the decision to participate in this particular technology is one of national sovereignty on the grounds that nations control the use of their e164 codes. RFC 3026, *supra* note 19, para 1, *at* ftp://ftp.rfc-editor.org/in-notes/rfc3026.txt. Ad Hoc ENUM Report, *supra* note 2, Sec. 4.1 (describing ENUM as an opt-in system for nations), *at* http://www.cybertelecom.org/library/enumreport.htm.

[25] *See* NeuStar FAQ, *supra* note 4, FAQ-8 (stating "Optimally, the root should contain a small listing of all of the national ENUM top-level country code name servers."), *at* http://www.enum.org/information/files/enum_faq.pdf.

[26] Brown, ENUM Service Reference Model, *supra* note 4, Sec. 4, *at* http://www.ietf.org/internet-drafts/draft-ietf-enum-operation-02.txt

[27] *See* Contribution of NeuStar, US Study Group A Ad-Hoc, *supra* note 4, p. 4.

[28] *See* Pfautz, ENUM Administrative Process, *supra* note 4, Sec. 1, *at* http://www.ietf.org/drafts/draft-pfautz-yu-enum-adm-01.txt.

[29] M. Mealling, R. Daniel, IETF RFC 2915, The Naming Authority Pointer (NAPTR) DNS Resource Record (Sept. 2000), *at* http://www.ietf.org/rfc/rfc2915.txt. *See* Ad Hoc ENUM Report, *supra* note 2, Sec. 2 (detailing use of NAPTR records), *at* http://www.cybertelecom.org/library/enumreport.htm.

[30] Jordyn A. Buchanan, Register.com, SGA Ad Hoc - ENUM, slide 5 (Feb. 12, 2001) (hereinafter Register.com, SGA Ad Hoc).

[31] *Id.,* slide 6-10.

This hierarchical model[35] creates an open platform where any service provider who receives authorization may participate. The full extent of what it means to be authorized and who issues the authorization is undefined and could impact on how open a system this model is. The database here would be unified and validated at Tier 1.

The rationale for this model is that it is based on the DNS and the DNS requires a single authoritative root for each node in the DNS tree.[36] If multiple roots existed, the question arises concerning how an ENUM device would know which database to look into and how an ENUM device could resolve inconsistent results from inconsistent databases. It is argued that a single root is required to ensure the integrity of ENUM.[37]

Alternative ENUM models suggest that ENUM can be provisioned as a wholly competitive service without need for a government sanctioned unified database. Detailed examination of the rational in favor of this argument will be visited in the Issues section below. In short, this contingent argues that ENUM is standardized data in an open database. Multiple ENUM services located in different domains therefore presents no significant challenge. On the occasion where the user does not know the full ENUM number, including its domain, the ENUM device can conduct a look up in all known ENUM services or the user could take advantage of a search engine. Once acquired, the information could be essentially "bookmarked" and search would not need to be repeated. Removing government regulation from this version of ENUM would make implementation faster, more flexible, and more responsive to consumers.

---

[32] *See* ICANN | Home Page (n.d.) *at* http://www.icann.org.

[33] D. Ranalli, D. Peek, R. Walter, IETF Informational Internet Draft, Tier-1 ENUM System Roles and Responsibilities, Sec. 4.4 (Feb. 2001) (hereinafter Ranalli, Tier-1 ENUM), *at* http://www.ietf.org/internet-drafts/draft-ranalli-peek-walter-enum-t1roles-01.txt.

[34] Brown, ENUM Service Reference Model, *supra* note 4, Sec. 4.1 (stating that "information changes infrequently"), *at* http://www.ietf.org/internet-drafts/draft-ietf-enum-operation-02.txt.

[35] The tiered model is detailed is multiple documents. *See* Ad Hoc ENUM Report, *supra* note 2, Sec. 5, *at* http://www.cybertelecom.org/library/enumreport.htm; Ranalli, Tier-1 ENUM, *supra* note 33, *at* http://www.ietf.org/internet-drafts/draft-ranalli-peek-walter-enum-t1roles-01.txt; Brown, ENUM Service Reference Model, *supra* note 4, Sec. 5, *at* http://www.ietf.org/internet-drafts/draft-ietf-enum-operation-02.txt; Pfautz, ENUM Administrative Process, *supra* note 4, Sec. 1, *at* http://www.ietf.org/drafts/draft-pfautz-yu-enum-adm-01.txt; Contribution of NeuStar, US Study Group A Ad-Hoc, *supra* note 4, p. 5; Register.com, SG-A Ad Hoc, *supra* note 30.

[36] *See* Contribution of NeuStar, Inc., US Study Group A Ad-Hoc, *supra* note 4, p. 3.

[37] *Id.,* p. 6.

### *Directory Services Market*

ENUM provides a directory service, providing a means of finding an individual through aggregated address information. The market for directory services is competitive. Competition comes from different services, different strategies, and different protocols.

## ENUM Projects

There are numerous ENUM projects. Some are essentially IETF ENUM implementations (marked by usage of a golden tree using a single top domain) and other alternative implementations.[38] All ENUM projects enter data in a standardized format into the open database DNS.

**NeuStar**: NeuStar is the current administrator of NANP. NeuStar, in a joint venture doing business as NeuLevel, was also recently awarded the new Top Level Domain (TLD) ".biz".[39] NeuStar has led the IETF effort, working with the ITU, and setting up the domain *E164.arpa*. NeuStar is operating an ENUM trial at *enum.org*.[40]

**I-TAB**: Jeff Pulver,[41] NetNumber, and I-Tab jointly applied to ICANN for the creation of the new TLD *.tel*.[42] This application was opposed by the ITU[43] and turned down by ICANN.[44] The Internet-Telephony Addressing Board was created as a part of the *.tel* application.[45] After the *.tel* application was denied, the I-TAB website went dark for a short period. It is back online with the stated mission of providing "an open industry forum for promoting the use of the ENUM standard by sharing operational experiences and by advancing operational recommendations for the delivery of ENUM based communications services."[46] The Board of Directors of I-TAB includes Jeff Pulver, Pulver.com, David P. Peek, NetNumber.com, Ike Elliot, Level3, Greg Vaudreuil, Lucent Technologies, and Jonathan Taylor, Voxeo.[47]

---

[38] US industry ENUM supporters acknowledge that there will be alternative ENUM implementations and recommend that such alternatives not be precluded. Ad Hoc ENUM Report, *supra* note 2, Secs. 1 & 4.1, *at* http://www.cybertelecom.org/library/enumreport.htm.

[39] NeuStar Press Release, NeuLevel Awarded Dot BIZ Top Level Domain by ICANN Board (Nov. 17, 2000) (stating "NeuLevel is a joint venture of NeuStar, Inc. and Australian based Melbourne IT, Ltd."), *at* http://www.neustar.com/pressroom/announcements/press_release.cfm?press_id=28.

[40] ENUM.ORG > Welcome Page (n.d.), *at* http://www.enum.org.

[41] Jeff Pulver is President and CEO of Pulver.com, Founder of the Voice on the Network Coalition, and a well known advocate for IP telephony. *See* The Jeff Pulver Homepage (n.d.) *at* http://www.pulver.com/jeff/.

[42] Jeff Pulver, David Peek of I-TAB, Glenn W. Marschel, NetNumber, TLD Application for .tel (Oct. 11, 2000), *at* http://www.icann.org/tlds/tel1/.

[43] Letter from ITU on Telephony-Related TLDs (Nov. 1, 2000) (hereinafter ITU Letter) (opposing ".tel" ENUM applications), *at* http://www.icann.org/tlds/correspondence/itu-response-01nov00.htm.

[44] *Net name body OKs seven new domains,* C|NET, (Nov 16, 2000), *at* http://news.cnet.com/news/0-1005-200-3730464.html.

**VeriSign**: VeriSign (a.k.a., NSI) and Telcordia partnered together to create ENUMWORLD. ENUMWORLD was created for the purpose of creating an ENUM testbed.[48] VeriSign announced in February 2001 the commercial launch of WEBNum. WEBNum is an ENUM-style service for use on wireless devices.[49]

**NetNumbers**: NetNumbers provides "secure, reliable, ENUM-compliant directory services to the Internet-Telephony industry."[50] NetNumber launched "the first ENUM directory service" in November of 2000, which is currently running without use of a government sanctioned golden tree. It owns the E164.com domain.[51] NetNumbers has created ENUM partnerships with Webley, Sonus Networks, Voxeo, Pingtel, SS8 Networks, Pagoo, Centile, OSPA, SIP Center.com, Nextone, Broadsoft, 2wire, Mediatrix, and Indigo.[52]

### Directory Services Competitors

ENUM also faces competition from other directory service projects. The first set described follows the strategy of aggregating multiple addresses into a single searchable database.

Dialnow permits subscribers to create a webpage containing their contact information. The data is accessed through the Dialnow.com database, using a telephone number as a query, on the dialnow.com webpage or through WAP devices. Dialnow claims that it has filed a patent on its technology, which may create a risk of future litigation with ENUM projects.[53]

---

[45] Internet Telephony Addressing Board, I-TAB (n.d.) ("The mission of the "Internet-Telephony Addressing Board" (ITAB) is to provide an open industry forum for promoting the use of the ENUM standard by sharing operational experiences and by advancing operational recommendations for the delivery of based communications services."), *at* http://www.i-tab.org.

[46] *Id*.

[47] I-TAB Home: Directors (n.d.), *at* http://www.i-tab.org/.

[48] *See* ENUM World Home (n.d.), *at* http://www.enumworld.com.

[49] VeriSign, Inc. - WEBNum (n.d.), *at* http://www.webnum.com.

[50] NetNumber Global ENUM Service (n.d.) *at* http://www.netnumber.com/.

[51] NSI - WHOIS Search Result: E164.COM (Jul. 19, 2001).

[52] NetNumber Global ENUM Service (n.d.), *at* http://www.netnumber.com/.

DotPHone (.ph) is the ccTLD of the Philippines.[54] A part of the ".ph" ccTLD is the dotPHone service. dotPHone provides users the opportunity to register domain names based on their name (instead of a telephone number). Users would then query the ".ph" name server with that domain name and receive the current addressing information for the registrant. If the user wanted to call John Doe, the user would enter the domain name John.Doe.ph and, if there were a record, receive the lasted telephone number.[55] In this way, it is almost identical to ENUM, with the exception that the single identifier appears to be a domain name of the registrants choice such as their name, as opposed to a public telephone number.

There are multiple Internet directory assistance projects. Essentially, online white pages or 411, these companies acquire subscriber list information pursuant to Sec. 222 of the Telecommunications Act[56] and upload the information as a searchable database. This is a highly competitive market that includes Switchboard, Anywho, Worldpages, 555-1212.com, MSN Reverse Look Up, Netscape White Pages Reverse Look Up, The Ultimate White Pages, Yahoo People Search, and Whowhere.

Unified Messaging also seems to be a service that follows the aggregation of addresses strategy to provide a unified means of reaching an individual.[57]

Microsoft recently announced its .NET Hailstorm project. Hailstorm's Passport user authentication system appears to be similar to ENUM in that it places a large amount of personal information behind a single means of accessing that information. It includes addressing information through such services as myAddress and myContacts. But Hailstorm has a wider versatility, usable for multiple types of interactions on the Internet with such services as myWallet, myProfile, and myCalendar.[58] In addition, Microsoft promises that the creator of such records will be able to control who has access to the records and how much of the records they have access to.

---

[53] DailNow.Com - The Internet Phone Company (n.d.), *at* http://www.dialnow.com/Investor_Information.asp.

[54] IANA Root-Zone Whois Information: .ph - Philippines (Jun. 18, 2000), *at* http://www.iana.org/root-whois/ph.htm.

[55] DotPHone, Both dotCOM domains and dotPH domains are functionally identical (n.d.), *at* http://www.domains.ph/answer.html.

[56] 47 U.S.C. § 222.

[57] *See*, e.g., Unified Messaging, E-mail, Fax, Voicemail, SMS, Phone all in one In-Box (n.d.), *at* http://www.unified-messaging.com.

[58] Microsoft, Building User-Centric Experiences: An Introduction to Microsoft Hailstorm (Mar. 2001), *at* http://www.microsoft.com/net/hailstorm.asp. *See also* ZDNet Onebox (n.d.) (offering voicemail, fax and email through one package), *at* http://www.zdnet.com/onebox/about.html.

Another competitor providing these types of services are Palm Pilots and similar address book software. There are current negotiations between phone manufacturers and Palm Pilot type device manufacturers concerning partnerships.[59] Wireless phones are being built with Palm Pilot type address books built in, giving ready access in the telephony device to known addresses of acquaintances. Having this information already in the phone could make ENUM services superfluous for most communications.

### Other Alternatives

There is a set of services that seeks to address the problem of how to find someone with a single address and building multiple communications applications on top of that address. If the user knows the single address, the user can use fax, telephony, messaging, or other applications to contact the desired individual at that address. This strategy is followed by SIP[60] and Instant Messaging.

In addition, the IETF's Telephony Routing Over IP (TRIP) protocol can be used to get telephony calls from the IP network to the PSTN. The protocol calls for the creation of a peer-to-peer network where participating servers announce available routes and gateways from an IP network to telephone numbers on the PSTN.[61]

### *What ENUM is Not*

ENUM is not an application. ENUM is a database. It is queried with an ENUM number and responds with contact data. Consequently, ENUM is not telephony. ENUM can be used is association with a multitude of applications including telephony, email,[62] fax, and others.[63]

---

[59] *See* Sprint PCS Press Release: Sprint PCS Phone QCP-6035 by Kyocera and Mobile Connectivity Kits for Palm Handhelds Are First in a Series of Palm Powered Solutions Offered By Sprint PCS (Apr. 11, 2001), *at* http://www.prnewswire.com/cgi-bin/micro_stories.pl?ACCT=153400&TICK=PALM&STORY=/www/story/04-19-2001/0001473179&EDATE=Apr+11,+2001.

[60] Tony Rutkowski, ENUM Directory Services in the Marketplace, DTI Workshop on ENUM, Slide 6 (Jun. 5, 2001) (noting "Email or SIP addresses may be more attractive.")

[61] J. Rosenberg, H. Salama, M. Squire, IETF Internet Draft, Telephony Routing over IP (TRIP) (Jun. 2001), *at* http://www.ietf.org/drafts/draft-ietf-iptel-trip-07.txt. *See also* Ad Hoc ENUM Report, *supra* note 2, Sec. 6.2.1 (noting role of TRIP), *at* http://www.cybertelecom.org/library/enumreport.htm.

[62] *See* Lind, CallFlows, *supra* note 4, para 2 (noting use of ENUM with email), *at* http://www.ietf.org/internet-drafts/draft-lind-enum-callflows-01.txt.

[63] *See supra* footnote 5 and accompanying text (noting different uses of ENUM).

ENUM does not do call set up.[64] The ENUM database provides data that the communication device may use to set up a call, but ENUM itself is more analogous to directory assistance.

ENUM is not a part of the public telephone network. ENUM does not interact with the SS7 network. An ENUM device is on the Internet, the ENUM query is over the Internet, and the ENUM database is a part of the Internet DNS database. Once the user obtains address information, the user may set up a call on the SS7 network, but that is separate and after the use of the ENUM protocol.

## Issues

ENUM is described as a convergence technology between the PSTN and the Internet world. This can make things messy. It may mean that policy considerations must consider the implications for both the regulated PSTN world and the unregulated Internet world. In this way, ENUM could be described more as a collision than convergence, bring both the best and the worst of both worlds together.

### *A Number by Any Other Name…*

Essential to ENUM is the connection of telephone numbers to ENUM numbers. This connection determines who has the right to assignment of an ENUM number and what government authority has jurisdiction over ENUM administration. If the connection is, however, broken, ENUM will be confronted with multiple challenging problems.

An ENUM number is a domain name. It could be anything that a domain name could be. The IETF ENUM Working Group was attempting to solve the problem of how to find devices on the Internet with two parameters. First, the IETF ENUM WG wanted to be able to do this using a numeric keypad. This limits an ENUM number to a numerical string. But it could still be any numerical string. Next, the IETF ENUM WG wanted to take advantage of phone numbers.[65] But the IETF ENUM WG could have select other types of numbers, as is demonstrated by current ENUM work considering the use of E.212 numbers with ENUM.[66] The IETF ENUM WG determined, by convention, an assignee of an ENUM number would use the same numerical string as the assignee's public telephone number.

---

[64] Brown, ENUM Service Reference Model, *supra* note 4, Sec. 4, *at* http://www.ietf.org/internet-drafts/draft-ietf-enum-operation-02.txt (stating "It is up to the client initiating the service request to sort through the set of NAPTR records to determine which services are appropriate for the intended action.")

[65] *See* footnote 10, and accompanying text.

[66] Gopal Dommety, Paddy Nallur, Viren Malaviya, Niranjan Segal, IETF Internet Draft, E.212 number and DNS (June 2001) (stating "This draft is adaptation of RFC 2916 to E.212 numbers."), *at* http://www.ietf.org/internet-drafts/draft-dommety-e212-dns-00.txt.

An ENUM number, however, is not itself a telephone number. A telephone number is an address used on the telephone network to reach a telephone.[67] An ENUM number is not an address. There is no communications device that is assigned and can be reached by using an ENUM number. You cannot set up a communications with an ENUM number itself. An ENUM number is a "token" used to query a database. This is the only function of an ENUM number. The database contains the addresses that can then be used in communications.

A numerical string standing by itself is a numerical string and is nothing more out of context. It becomes a type of number in a particular context. 5550100 is a numerical string. Use this number to reach a telephone on the telephone network and it is a telephone number. Use this number to access money in a bank account and it is a bank account number. Use this number to access an ATM and it is a PIN. What type of number a numerical string is, depends upon the context in which it is used. Outside of that context, it is no longer that type of number. Simply because two numerical strings have the same value does not make them the same type of number.

Good examples are other databases tied to telephone numbers such as grocery store savings plans and video rental membership. If you forget your card you can give the cashier your phone number and you have access to the relevant the database. The mere use of a phone number in a database does not give the FCC jurisdiction over grocery store savings plans or video clubs.[68] The reason why is, in that given context, the numerical string has the same value as a telephone number but is, in fact, a savings plan number. The use of the telephone number serves as a pneumonic device but has no further connection to the telephone network. There is a difference between something *being a* telephone number and *having the same value as* (same numerical string) a telephone number.

Members of the ENUM industry implicitly recognize this point. Documents that describe ENUM discuss it as transferring one number into another number. The industry repeatedly uses such works as mapping,[69] tied,[70] translating,[71]

---

[67] Federal Standard 1037C, Definition: telephone number (Aug. 23, 1996) (stating "telephone number: The unique network address that is assigned to a telephone user, i.e., subscriber, for routing telephone calls."), *at* http://glossary.its.bldrdoc.gov/fs-1037/dir-036/_5369.htm.

[68] While it is true that the ENUM database is unlike the others cited in that the ENUM database contains communications data, it is also true that a great deal of that communications data is data that the FCC lacks jurisdiction over, including e-mail addresses, web addresses, IP telephony addresses, physical addresses, and other personal identifying information. To suggest that the FCC has jurisdiction just because a phone number is in the database would also be to suggest that the US Post Office would have jurisdiction over ENUM because the database would likely contain physical addresses as well.

[69] Brown, ENUM Service Reference Model, supra note 4, Sec. 5.1, *at* http://www.ietf.org/internet-drafts/draft-ietf-enum-operation-02.txt; *SS8 Links Multiple PSTN and IP Devices to Single Phone Number*, COMMUNICATIONS DAILY, p. 7, Jun. 25, 2001.

*Robert Cannon*  *Draft* - *Page 15*

transforming,[72] and converting[73] to describe this process. ENUM is also described as a "telephone number-*based* Internet directory service."[74] All of this recognizes the process of taking one numerical string out of its original context and using it in a new context.

The argument that ENUM numbers and telephone numbers are distinct is supported by the fact that the two types of numbers are operationally distinct. Telephone numbers can operate without ENUM; telephone numbers can cease to operate regardless of ENUM. ENUM numbers, which can be anything, can technically be created without a corresponding telephone number. An ENUM number can be deleted from the DNS without an affect on the telephone number. Telephone numbers are used on the telephone network; ENUM numbers are used on the Internet.

This is highlighted by one of the primary issues for ENUM: what happens when a telephone number is disconnected? The ENUM industry is working hard on developing relationships so that ENUM service providers can be informed when a telephone number is terminated.[75] If the numbers were the same, then when a telephone number ceased to exist, the ENUM number could no longer function. The fact that the ENUM number can technically live on when no corresponding telephone number is in existence demonstrates that they are distinct. The connection between telephone numbers and ENUM numbers has to be established by convention because it is not established by law or technical requirement.

The reasons why the distinction is important are jurisdiction, authority, and rights to a number. If ENUM numbers are telephone numbers, then they possibly fall under the jurisdiction of telephone authorities. If, however, ENUM numbers are not telephone numbers, then they do not necessarily fall under the jurisdiction of telephone authorities. In addition, there would be no right to an ENUM number based on being the assignee of a telephone number. This could complicate conflicts over ENUM number assignments and who has authority to set up Tier 1 ENUM providers.

---

[70] Pfautz, ENUM Administrative Process, *supra* note 4, Sec. 1, *at* http://www.ietf.org/drafts/draft-pfautz-yu-enum-adm-01.txt; Penn Pfautz, ENUM Administration, Slide 2 (Feb. 12, 2001), *at* http://www.itu.int/infocom/enum/workshopusafeb12-13/pfautz.htm.

[71] Shockey, SGA, *supra* note 10, slide 2; Marc Robins, *ENUM's Got Your Number*, INTERNET TELEPHONY, Jun. 2001, *at* http://www.tmcnet.com/it/0601/0601ms.htm.

[72] Ad Hoc ENUM Report, *supra* note 2, Sec. 2, *at* http://www.cybertelecom.org/library/enumreport.htm.

[73] *SS8 Links Multiple PSTN and IP Devices to Single Phone Number*, COMMUNICATIONS DAILY, p. 7, Jun. 25, 2001.

[74] Brown, ENUM Service Reference Model, *supra* note 4, Secs. 3, 4 (emphasis added), *at* http://www.ietf.org/internet-drafts/draft-ietf-enum-operation-02.txt.

[75] Ranalli, Tier-1 ENUM, *supra* note 33, Secs. 4.2, 6.3.

This is an intriguing issue of rights to numbers. Rights to one type of number do not transfer to another type of number simply because the numerical strings are the same. Otherwise, rules and regulations concerning one type of number in one context developed with a particular history and concerns, would be applied to foreign numerical strings and in alien contexts. The rules and regulations of one situation would be expanded to reach contexts never anticipated or intended. Well founded restrictions on one type of number could be irrational in another context. An individual with one type of number could control the use of that numerical string in other contexts, extracting fees or concessions for its use. This could create a dangerous precedent and have far reaching ramifications.

### *DNS Issues*

The core issues raised by ENUM are issues of administration the DNS database. The core issue for a national government to resolve is whether to sanction a national Tier 1 service provider and related administrative issues.[76]

#### **Unified Database**

The first issue raised is whether ENUM requires a unified global database, also known as a "global tree." Proponents of a unified database argue that if there are multiple databases, an ENUM device would not know which to query. Furthermore, there is a risk of incompatible records in different databases.[77]

Even if it is assume that a unified database is needed, one already exists. ENUM is a DNS innovation and the DNS is a unified database. Any user anywhere in the world can query a DNS name server for www.cybertelecom.org and they will get the appropriate result. The DNS is both unified and global. Thus, the question presented by ENUM is whether there needs to be a unified database *inside* the unified database of DNS.

Pursuant to the ENUM protocol, data would be entered into the open DNS in a standardized format. Since the data exists in a standard format across open, interconnected, distributed databases, searches of that data are relatively easy. If there were multiple ENUM databases, and if a user did not know which one to search, an opportunity would be created for metasearch engines to be created, creating an ability to find the data in any known database. Alternatively, an ENUM resolver could query known ENUM databases to determine if records

---

[76] Pfautz, ENUM Administrative Process, *supra* note 4, Sec. 1, *at* http://www.ietf.org/drafts/draft-pfautz-yu-enum-adm-01.txt

[77] Ad Hoc ENUM Report, *supra* note 2, Sec. 4.1, *at* http://www.cybertelecom.org/library/enumreport.htm.

*Robert Cannon*           *Draft - Page 17*

exist.[78] NetNumbers indicates that it already has such a publicly available resolver.[79]

As consumers could access the information in the open DNS at multiple ENUM service providers as easily as a single provider, there is nothing that would drive the consumers to use only a single provider. Network effect is a factor for ENUM as a whole (for ENUM to work there has to be overall network effect), but not for individual competitors. In other words, if ACME ENUM has only a few thousand records, but is reachable through metasearch engines, a resolver, or the use of extensions, then ACME could have as competitive a place in the market as large service providers.

In addition, if the issue with multiple databases is knowing which database to search, the answer would seem obvious: tell the ENUM device which database to search. One possible way in which this could be achieved is by adding extensions to numbers. 5551212#36 could mean NetNumbers where 5551212#46 could mean NeuStar. Since the device now knows which database to look in, this is no longer an issue.[80]

Furthermore, ENUM databases, due to network effect, have an incentive to cooperate. ENUM has more value if it has more data; a means of getting more data is to cooperate with other ventures and create open data platforms.[81]

While it is not clear that a Golden Tree approach is necessary[82], such an approach could have advantages. A centralized database could arguably facilitate data verification, authentication, and integrity. Through a central database, only data that met specifications would be entered. Unverified data would be rejected and only one record for a given number would be created. Competitive service providers would be interconnected through the unified database.

---

[78] Ad Hoc ENUM Report, *supra* note 2, Sec. 4.1 (noting alternatives to golden tree approach), *at* http://www.cybertelecom.org/library/enumreport.htm.

[79] Douglas Ranalli, Is E164.arpa The Only Answer for Tier-1 ENUM Registry Services? (n.d.) (also noting that "there is no evidence of the market deployment of hundreds or thousands of ENUM services," meaning that querying those ENUM services that exist would be manageable), *at* http://www.netnumber.com/news/e164arpaComp.pdf.

[80] ENUM also seeks to solve the problem of telephone restrained by merely having numeric keypads with which to enter addresses. New wireless phones have touch screens that can be configured in any way for any type of data input, increasing the opportunity for address design and ability to designate the appropriate database. *See* Kyocera - Kyocera SmartPhone Series (n.d.) (showing wireless phone with touch screen in place of keypad), *at* http://www.kyocera-wireless.com/kysmart/kysmart_series.htm.

[81] Ad Hoc ENUM Report, *supra* note 2, Sec. 4.1 (noting possible interconnection alternative to golden tree approach), *at* http://www.cybertelecom.org/library/enumreport.htm.

[82] *See also* Ad Hoc ENUM Report, *supra* note 2, Sec. 8.1 (Minority View of Report, indicating alternative to golden tree implementation), *at* http://www.cybertelecom.org/library/enumreport.htm

Additionally, a joint partnership could have the advantage of branding and joint marketing. A joint effort can be marketed to the public as the service endorsed widely by industry participants. [83]

A disadvantage of a global unified database is the tremendous amount of global coordination required in order to succeed. There could be 150+ Tier-1 service providers that need to be established and coordinated. The effort involved in order to achieve coordination may result in delay in ENUM implementation and administration.[84] An additional disadvantage is possible restraints on creativity and innovation. As ENUM is administered is highly centralized through a global system, innovations could only be achieved through that centralized structure. This reduces the ability of a competitive process to create new solutions that users might desire.[85]

Whether the Golden Tree approach is adopted may not immediately rise to a public policy concern if further questions are not reached. In other words, if a Golden Tree does not require government sanction, then numerous concerns are alleviated. However, if industry continues to press for a government sanctioned Tier 1 provider, it must be recognized that the election of the Golden Tree approach is one of preference and not necessity. In other words, selecting a unified approach which requires regulatory intervention and the creation of a government sanction monopoly is a path of choice and it could be avoided.

### E164.arpa?

If it is concluded that there should be a unified database, where should that database be located?[86] RFC 2916 indicates that IANA should delegate the domain name *e164.arpa* pursuant to the recommendation of Internet Architecture Board (IAB).[87] Pursuant to IAB recommendation, *e164.arpa* is to be technically administered by RIPE NCC.[88] The IETF selected *e164.arpa* as the location of the ENUM database because *.arpa* is dedicated to infrastructure issues and is well managed, stable and secure.[89]

---

[83] A concern has been raised that if multiple ENUM service providers form a joint partnership to implement ENUM, there could be antitrust concerns. It is beyond the scope of this paper to do a proper antitrust analysis.

[84] *See* Doug Ranalli, Is E164.ARPA The Only Answer For Tier-1 ENUM Registry Services? (n.d.) (noting delay resulting from global coordination), *at* http://www.netnumber.com/news/e164arpaComp.pdf

[85] *See Id.* (noting impact on creative process).

[86] If it is concluded that a unified database is not needed, then there is no reason to reach the question of whether it should be located at *e164.arpa* or elsewhere.

[87] RFC 2916, *supra* note 2, Sec. 4 (stating "This memo requests that the IANA delegate the E164.ARPA domain following instructions to be provided by the IAB."), *at* http://www.ietf.org/rfc/rfc2916.txt.

[88] E164.ARPA WHOIS Record, Network Solutions (May 17, 2001). While IAB Meeting minutes reference *E164.arpa*, no record of IAB instruction to IANA for delegate to RIPE NCC has been found.

France has objected to this arrangement and argued that the administration should be done under *e164.int* under ITU authority. France argued that management of ENUM must be subordinate to E.164 management, and that E164 management is under the authority of the ITU. Thus, the French argue that "the most coherent approach is obviously to use a suffix managed by the ITU."[90]

Robert Shaw of the ITU has argued that the ENUM DNS name servers need to be "dispersed around the world." He then points out that 8 of the 9 *.arpa* name servers are deployed in the United States and are not dispersed around the world.[91]

Originally *.arpa* was the domain of the US Defense Advanced Research Projects Agency (DARPA). On April 14, 2000, DARPA disassociated itself with the *.arpa* domain with the understanding that *.arpa* would be dedicated to infrastructure (along with *.int*) under the authority of the Internet Assigned Number Authority (IANA),[92] which is currently a part of ICANN.[93] There was an effort to rename

---

[89] Ad Hoc ENUM Report, *supra* note 2, Sec. 8.1, *at* http://www.cybertelecom.org/library/enumreport.htm; Shockey, ITU-T, *supra* note 4, slide 11; Shockey, SGA, *supra* note 10, slide 13.

[90] France, Conditions for Implementation of ENUM, ITU SG2 Delayed Contribution on D.15-E (Jan. 23, 2001), *at* http://www.ngi.org/enum/pub/15_ww9.htm.

[91] Shaw, DTI ENUM Workshop, *supra* note 11, slides 13-15, at http://www.itu.int/infocom/enum/dtijune501/dti-june-5-2001-1.PPT. *Compare* Shaw, ICANN (where this argument appears to have been omitted), *at* http://www.itu.int/infocom/enum/GACjune1201/gac-june-2-2001-1.PPT.

[92] *.arpa* and *.int* are designated as Internet infrastructure domains to be managed by IANA. *See* Jon Postel, IETF Draft, New Registries and the Delegation of International Top Level Domains, para 1.3 (May 1996) (stating that *.arpa* and *.int* were "created for technical needs internal to the operation of the Internet at the discretion of the IANA in consultation with the IETF."); IAB Statement on Infrastructure Domain and Subdomains (May 10, 2000), *at* http://www.iab.org/iab/DOCUMENTS/statement-on-infrastructure-domains.txt; Annex 8: Responsibilities for e164.arpa, Sec. (2) (n.d.) ("IAB requested on May 17 2000 that assignment of subdomains of arpa should be a task of IANA."), *at* http://www.itu.int/infocom/enum/workshopjan01/annex8-responsibilitiesfore164.arpa.htm; IANA | Contact Information (modified November 3, 2000), *at* http://www.iana.org/contact.htm; Letter from Karen Rose, NTIA Purchase Order Technical Representative, to Mr. Louis Touton, Vice-President, Secretary, and General Counsel, ICANN (Apr. 28, 2000) (hereinafter Rose Letter) ("The Department of Commerce considers this an Internet Assigned Numbers Authority (IANA) function and has requested that the WHOIS entry for the ARPA domain reflect IANA as the registrant."), *at* http://www.ngi.org/enum/pub/DOC_28Apr2000.htm.

[93] *See* Contract Between ICANN and the United States Government for Performance of the IANA Function (Feb. 9, 2000), *at* http://www.icann.org/general/iana-contract-09feb00.htm; IETF Informational RFC 2860, Memorandum of Understanding Concerning the Technical Work of the Internet Assigned Numbers Authority (Jun. 2000), *at* ftp://ftp.ietf.org/rfc/rfc2860.txt; Rose Letter, *supra* note 92 (stating "Purchase Order No. 40SBNT067020 provides that '[ICANN] will perform

ARPA domain as the *Address and Routing Parameter Area* in an attempt to distinguish it from US DARPA.[94] IANA administers *.arpa* in compliance with IETF protocols.[95] *.arpa* has been traditionally used for reverse-DNS lookup.[96] US industry notes that *.arpa*, unlike *.int*, meets the security, performance, and reliability requirements of an infrastructure domain as set forth in IETF RFC 2870.[97]

*.int* was originally a infrastructure domain along with *.arpa*.[98] Currently it is dedicated to international treaty organizations.[99] *.int* is not under the control of the ITU.[100] Placing ENUM under *.int* does not necessarily place it under the control of the ITU or anyone else.

The selection of TLD itself may not be significant. The most compelling argument in favor of *.arpa* is that the infrastructure related to it is superior. But the infrastructure related to *.int* could be upgraded if necessary (assuming someone bore the cost). Perhaps the most compelling difference is one of appearance. If ENUM is under *.int*, there is an appearance that it is under greater ITU control. If it is under *.arpa*, there is an appearance that it is under greater IETF control. But under ENUM as currently envisioned, the user will be aware of the ENUM number, not the TLD. In the final analysis, this issue may be one of sound and fury, signifying very little.

---

other IANA functions as needed upon request of DOC.'"), *at* http://www.ngi.org/enum/pub/DOC_28Apr2000.htm.

[94] Rose Letter, *supra* note 92, *at* http://www.ngi.org/enum/pub/DOC_28Apr2000.htm.

[95] B. Carpentar, F. Baker, M. Roberts, IETF Informational RFC 2860, Memorandum of Understanding Concerning the Technical Work of the Internet Assigned Numbers Authority, Sec. 4 (June 2000) (indicating that disputes between IANA and IETF are resolved by IAB), *at* http://www.ietf.org/rfc/rfc2860.txt.

[96] *See* IETF Best Current Practice RFC 2317 Classless IN-ADDR.ARPA delegation (March 1998), *at* http://www.ietf.org/rfc/rfc2317.txt. in-addr.arpa domain "is used to convert 32-bit numeric IP addresses back into domain names. This is used, for example, by Internet web servers, which receive connections from IP addresses and wish to obtain domain names to record in log files." Connected: An Internet Encyclopedia: The in-addr.arpa Domain (n.d.), *at* http://www.freesoft.org/CIE/Course/Section2/15.htm.

[97] Ad Hoc ENUM Report, *supra* note 2, *at* http://www.cybertelecom.org/library/enumreport.htm

[98] See discussion, footnote 92.

[99] J. Postel, IETF RFC 1591, Domain Name System Structure and Delegation, Sec. 2 (Mar. 1994), *at* http://www.isi.edu/in-notes/rfc1591.txt.

[100] However, there are some indications that the ITU is attempting to gain control of *.int*. *See* Joakim Stralmark, ENUM- functions that maps telephone numbers to Internet based addresses, Post & Telestyrelsen, 3 (Mar. 23 2001), *at* http://www.enum.org/information/files/enum_summary.pdf (stating "ITU has ambition of becoming the registrar for the top-level domain .int."); ITU, INT Top Level Domain Name Registration Services (January 15, 1999), *at* http://www.itu.int/net/int/.

## Government Sanctioned Monopoly?

If there is to be a unified database, how will it be administered and does it require a government sanctioned monopoly? The IETF ENUM model calls for ITU involvement at Tier 0 and national governments setting up Tier 1 providers. Even if it is assumed that Tier 0 and Tier 1 providers are necessary, government sanctioning of these providers would be inappropriate.

The possible benefits of creating a government sanctioned monopoly must be weighted against the costs. Such monopolies impact competition in their market; normally they eliminated competition in their market. This, in turn, has an impact on innovation and responding to consumer needs. The monopoly service becomes encumbered with government entanglement, dramatically reducing the speed of deployment and innovation. Centralized decision making in compliance with federal administrative law is slow and less responsive to needs. In addition, there is the cost of the bureaucracy and the lawyers and lobbyists employed to interact with that bureaucracy.[101]

Particularly problematic is the potential delay resulting from government involvement.[102] In order to implement a U.S. government sanctioned ENUM service, there must be (1) legislative authority, (2) regulation, and (3) a government procurement process. This could result is multiple years of delay in which alternatives could make the government sanctioned ENUM implementation obsolete. In addition, further evolution in ENUM policy would likewise be encumbered by government process.

At the international level, NetNumbers points out that "it is simply time consuming and difficult to coordinate the selection of Tier-1 ENUM service providers access 200+ ITU Member States."[103] The resources dedicated to "achieving consistent policies regarding registration procedures, conflict resolution, disclosure of registrant information, etc."[104] may significantly impede progress of ENUM in the International arena.

The issue of the delay caused by the need for government involvement may be one of the most insurmountable problems for ENUM.

---

[101] *See* Douglas Ranalli, *Is "E164.arpa" The Only Answer for Tier-1 ENUM Registry Services?* (n.d.) (stating that coordination at international level would be time consumer, difficult, and artificially limit creative process).

[102] *See* Anthony Ruthkowski, *the ENUM golden tree*, INFO (Apr. 2001) (recounting failed experience of standard X.500).

[103] Douglas Ranalli, *Is "E164.arpa" The Only Answer for Tier-1 ENUM Registry Services?* (n.d.).

[104] *Id.*

### Technological Viability

ENUM is not a final IETF standard; it is a proposed standard.[105] A proposed standard is a standard on paper that has not been tested or tried. Although it is a stable standard, it is subject to change based on further experience. An RFC becomes a final "Internet Standard" when it has a significant implementation, is operationally successful, and has a "high degree of technical maturity."[106] ENUM, as of yet, has not demonstrated that it is a mature technology. Government sanctioning of a standard that is not final would be unusual.

### Commercial Viability

Whether ENUM is likely to be commercial viable is less then certain. There are no known consumer studies concerning whether ENUM is a service that consumers desire. There has been limited trial market deployments.[107] Even if ENUM were to be viable, there has been no study on what the market penetration might be (would it be widely deployed or useful only to a limited niche market) or whether the viability might be short lived.

Conversely, there are several indicators that suggest that ENUM may have difficulty being commercially viable. The primary concern is privacy; people may not want all of their contact information aggregated in a single open space. Similarly, ENUM is mono dimensional; an ENUM number goes in and all of the contact information comes out, without flexibility or further alternatives. Alternatives, such as the proposed Microsoft Hailstorm offers greater consumer empowerment, offering greater control over what information will be released to different queries of the system. Based on privacy concerns, alternatives could be more compelling then ENUM's rigid option.

The second factor is network effect; ENUM will not be valuable unless a large number of individuals register ENUM numbers. But until there is a large number of registrations, there were be a low incentive to register with ENUM (a catch-22). Likewise, the numerous competitors to ENUM challenge its possibility for success. Even if ENUM enjoys a degree of success, it is unclear whether it will continue to enjoy such success. Telephones are becoming increasingly intelligent; ENUM's restraints, such as the limitation to the numeric keypad, may make it antiquated.[108] There is a possibility that ENUM seeks to solve yesterday's problem.

---

[105] *See* RFC 2026, *supra* note 1, (explaining IETF process and difference between proposed, draft, and Internet standards), *at* http://www.ietf.org/rfc/rfc2026.txt.

[106] *Id.*, Sec. 4.1.3.

[107] *See* Tony Rutkowski, ENUM Policy Briefing to US Dept of State, FCC, and NTIA, slide 9 (n.d.) *at* http://www.enumworld.com/resources/NTIA_policy_brief.ppt. NetNumbers is a live market deployment but not data has been presented from NetNumbers indicating success of the deployment.

[108] *See* footnote 80 (noting that modern phones offer greater flexibility for address input and need not be limited to numeric strings).

Further difficulty could be experienced internationally, where several countries have expressed concern over IP telephony bypass of the public telephone network and sought to bar such bypass. As ENUM could be perceived as facilitating bypass, it could be expected that several countries might bar ENUM, limiting its network effect and thus commercial viability.

The commercial viability of ENUM is not established and may even be doubtful. It would therefore be imprudent for a government to sanction a monopoly for a service where its viability is in question.

### Directory Assistance Competition

ENUM is a directory assistance service. It provides a solution to the problem of how to find a means of communicating with an individual. As noted above, the directory assistance market is highly competitive. ENUM faces competition from such powerful market players as Microsoft, AOL, VeriSign, and Palm Pilot.[109] A golden tree approach to ENUM would likely have to compete with private implementations of ENUM[110] (NetNumbers has been commercially launched since November of 2000 and has acquired 14 partners without any need of government sanctioning[111]). ENUM also faces competition from SIP, Instant Messaging, and TRIP. This competitive market gives users the ability to sort out which services are the most useful and compelling. Endorsement by the government of one competitor over all others would distort the market, be inappropriate, and determine market winners through regulation instead of competition.

### ITU Involvement

IETF presentations have indicated that all countries must address the same issues for ENUM.[112] There is no further explanation of why this is so. Given the wide diversity of regulatory and market environments, it would seem that any requirement that national tier 1 providers address ENUM issues in exactly the same way would be unnecessary, inaccurate, and cause significant delay while coordination is resolved.

---

[109] *See*, *supra* p. 10 (listing competitive alternatives to ENUM).

[110] Two documents so far have suggested that alternative implementations of ENUM should be restricted or prohibited. *See* France Conditions, *supra* note 90, *at* http://www.ngi.org/enum/pub/15_ww9.htm; Stralmark, *supra* note 100, *at* http://www.enum.org/information/files/enum_summary.pdf.

[111] *See* footnote 50, and accompanying text.

[112] See Steve Lind, AT&T, Tony Holmes, BT, ENUM Administration Issues, slide 5 (Jan. 17, 2001) (hereinafter Lind, ENUM Administrative Issues); Chairman's Report of the ITU ENUM Workshop, ITU, Geneva (Jan. 17, 2001), Annex 7: ENUM Issues: Issue 3, *at* http://www.itu.int/infocom/enum/workshopjan01/report-jan17-2001.htm

The IETF is cooperating with the ITU partly because the ITU is the authority for the E.164 numbering system. Originally, as stated in the ENUM RFC, the role of the ITU was limited:
> Names within this zone are to be delegated to parties according to the ITU recommendation E.164. The names allocated should be hierarchic in accordance with ITU Recommendation E.164, and the codes should assigned in accordance with that Recommendation.[113]

The role was limited to the fact that country codes in *e164.arpa* are to comport with the ITU E.164 Recommendation. The ITU had no authority pursuant to this text; it was not asked to do anything.

In October 2000, the ITU released the *Liaison to IETF/ISOC on ENUM*.[114] This Liaison requires national governments to designate to the ITU their Tier 1 service provider. Thus the ITU would act as an international ENUM gate keeper and credential recognizer. The Liaison also appears to attempts to obligate any ENUM effort, whether part of the golden tree or not, to comply with ITU direction.[115]

In June of 2001, Robert Shaw recommended an even further role for the ITU, suggesting that the ITU should be responsible for outsourcing the responsibilities of administering the Tier 0 service provider and "define and implement administrative procedures that coordinate delegations of E.164 numbering resources into these name servers."[116]

---

[113] RFC 2916, *supra* note 2, para 4, *at* http://www.ietf.org/rfc/rfc2916.txt

[114] Liaison, *supra* note 19, *at* http://www.itu.int/infocom/enum/wp1-39_rev1.htm. This was subsequently released as an informational RFC. RFC 3026, supra note 19, *at* ftp://ftp.rfc-editor.org/in-notes/rfc3026.txt. "An 'Informational' specification is published for the general information of the Internet community, and does not represent an Internet community consensus or recommendation." RFC 2026, supra note 1, para 4.2.2, *at* http://www.ietf.org/rfc/rfc2026.txt.

[115] According to the Liaison, "All administrative entities, including DNS administrators, will adhere to all the applicable tenets of all pertinent ITU Recommendations, e.g., E.164, E.164.1, E.190, and E.195, with regard to the inclusion of the E.164 resource information in the DNS." Liaison, *supra* note 19, *at* http://www.itu.int/infocom/enum/wp1-39_rev1.htm. The ITU's role is further described as follows: "For all E.164 Country Code Zone resources (Country Codes and Identification Codes), the ITU has the responsibility to provide assignment information to DNS administrators, for performing the administrative function. The ITU will ensure that each Member State has authorized the inclusion of their Country Code information for input to the DNS. For resources that are spare or designated as test codes there will normally be no entry in the DNS. However, the ITU will provide spare code lists to DNS administrators for purposes of clarification. The entity to which E.164 test codes have been assigned will be responsible for providing any appropriate assignment information to DNS administrators." *Id.* And again, "The ITU may request the consultation of the WP1/2 experts as necessary and as prescribed in Resolution 20." *Id. See also* Shockey, SGA, supra note 10, slide 18 (stating "ITU will insure that Member States have authorized inclusion of their Country Code in e164.arpa" and "ITU to coordinate with RIPE NCC as the Root Administrator."), *at* http://www.itu.int/infocom/enum/workshopusafeb12-13/shockey.htm.

[116] Shaw, DTI ENUM Workshop, supra note 11, slide 16 at http://www.itu.int/infocom/enum/dtijune501/dti-june-5-2001-1.PPT; Shaw, ICANN, *supra* note 11, slide 14, *at* http://www.itu.int/infocom/enum/GACjune1201/gac-june-2-2001-1.PPT.

One explanation for ITU involvement is the concern on the part of the IETF and RIPE that it does not want to be put into the position of determining who is the appropriate authority for an e164 code. If the ITU recognizes the credentials of an entity as the proper authority for that code, that relieves the IETF and RIPE of the risk of getting involved in skirmishes over who the proper authorities are.

The *ITU Liaison* design does not appear necessary. As articulated in RFC 2916, ENUM requires receiving the data of what E.164 country codes map to what countries. Other than this public available information that does not require ITU action or authority, there appears to be no need for ITU authority or involvement.

Much of the ITU's involvement is based on the premise that ENUM are telephone numbers, and the ITU is the authority over the E.164 standard. As demonstrated above, ENUM numbers are not telephone numbers.

The benefit of the RIPE NCC acquiring a gatekeeper must be weighed against the costs. There are other means by which this can be achieved. RIPE NCC could set forth the criteria for the representatives it will recognize. For example, RIPE NCC could indicate that the head of a nation's ITU delegation must specify the Tier 1 ENUM provider to RIPE NCC. The nation would interact directly with RIPE NCC without the ITU intermediary.

The relationship between the IETF and ITU is one of mutual recognition. The ITU Liaison recognizes the IETF effort and the IETF in turn recognizes ITU authority. By such recognition, the IETF ENUM effort is set apart from other private ENUM projects. Indeed, the ITU has opposed ENUM efforts that do not recognize the need for the ITU.[117] Mutual recognition is an insufficient justification for ITU authority and has a negative impact on competition.

### Joint Venture

If governments do not sanction ENUM service providers, the ENUM industry itself could cooperate and set up a unified tree ENUM project without the government. This could, for example, be a joint venture.[118] However, one concern with such cooperation would be anti trust concerns. NeuStar has cursorily concluded that there is no anti trust concern.[119] An antitrust analysis is beyond the scope of this paper. However, it is worth noting that the issue exists.

---

[117] The ITU sent a letter to ICANN opposing Pulver's application to create a new TLD *.tel*. ITU Letter, *supra* note 43, *at* http://www.icann.org/tlds/correspondence/itu-response-01nov00.htm.

[118] The ENUM industry seems to have implicitly recognized that it can set up a domestic ENUM golden tree without government involvement. During the summer of 2001 AT&T and WorldCom had competing proposals concerning how industry could cooperatively and without government involvement, launch ENUM domestically. Steven D. Lind, AT&T, U.S. ENUM Frame Document Implementation Framework (n.d.) (distributed at June 18, 2001 State Department ENUM Ad Hoc Meeting); Peter Guggina, WorldCom Contribution for Independent ENUM Forum (Jun. 12, 2001). *See also* Ad Hoc ENUM Report, *supra* note 2, Sec. 8.1 (discussing industry forum), *at* http://www.cybertelecom.org/library/enumreport.htm

### Conclusion

The question of whether ENUM should have government sanctioned monopoly providers is in the historical context of the deregulatory environment of the Telecommunications Act of 1996, the efforts to privatize the DNS through the work of ICANN, and the US's policy position that Internet issues are outside the jurisdiction of ICANN. The ENUM question runs directly into US policy in the area of IP Telephony and ICAIS[120] where the US has defending the notion countries will experience the greatest benefit from high tech innovation if they leave these markets unregulated. In an age where the government in embarked in a tireless battle to tear down monopoly positions in the market, ENUM asks that it be blessed with monopoly status.

Historically, the government sets up two types of monopolies: production monopolies or standards monopolies. Production monopolies are typified by AT&T in the 1930s where, in the opinion of the government, there was an efficiency in only have one company produce the service.[121] Standards require government sanction where there is something about the standard that compels sanctioning. The North American Numbering Plan (NANP) is a standard that requires unique assignment of telephone numbers. There can be only one.

ENUM fits within neither of these situations. The directory assistance market is competitive. The barrier to entry is low and is the risk monopolization. Conversely, sanctioning one competitor over others could thwart innovation and service to the consumer. Likewise, ENUM is not a standard that requires government sanction.

The ENUM industry has already made contingency plans, in the event that the US government fails to act, to implement ENUM domestically through an ENUM forum. They have conceded that government sanctioning is not necessary to make this succeed. The cost of having the government involved will like be multiple years of delay, giving alternatives first mover advantage and making that delay fatal to ENUM. Not only is government sanctioning of ENUM

---

[119] Contribution of NeuStar, US Study Group A Ad-Hoc, *supra* note 4, p. 2 (stating in one sentence and without supporting analysis that there is no antitrust concern).

[120] International Internet Carriage and Settlement or ICAIS. This is a controversy over Internet backbone peering, where other countries seek to impose telecommunications settlements on Internet peering and the US has opposed such policy in favor of unregulated private contractual negotiations in a competitive market. Material on ICAIS can be found at WIP: International (modified August 8, 2001), *at* http://www.cybertelecom.org/international.htm#icais.

[121] Milton Mueller, *Universal Service in Telephone History: a reconstruction*, TELECOMMUNICATIONS POLICY 17, 5 (July 1993) 352-69.

inappropriate, it would also probably assure that ENUM would never be a commercial success.[122]

### International Administration

If there is to be a government sanctioned unified database, then policy considerations about how that will be implemented will need to be considered. Internationally, the administrative contact for the Tier 0 provider at e*164.arpa* domain is the IAB and the technical contact is RIPE NCC. But the authority of the IAB and RIPE NCC is not clear. At the national levels, the Tier 1 service provider would have authority derived from the nation government. RIPE NCC and the IAB, however, have no international or national authority. This raises questions such as

- From where is their authority derived?
- To whom are they accountable?
- How will their ENUM work be funded?
- How would disputes be resolved?
- How would they behave in the event of war or national disaster?
- How would they be protected from litigation or local process (i.e., search warrants or wiretaps)?
- How would they be open and transparent?
- How would they be responsive to member states?
- How would they resolve new policy questions? Who would have the authority to resolve those questions?
- How will reliability be assured?[123]

Unless the authority for the Tier 0 provider is properly established, it could make ENUM vulnerable to continuous challenges and problems. It may be appropriate to consider whether the documentation behind *.arpa* and the delegation of *E164.arpa* to IAB and RIPE NCC is sufficient to be legally stable. If ENUM becomes essential to communications, it would be in the public interest to ensure its full stability and reliability.

### DNS Conflict Resolution

How will potential conflicts between ENUM numbers be resolved? In the DNS, ICANN regulates by contract, requiring domain name registrants to agree to be bound by the Uniform Dispute Resolution Process before WIPO. NeuStar has suggested that one appropriate solution for ENUM is compliance with ICANN's Uniform Dispute Resolution Process.[124] As ENUM numbers are domain names, it is possible that this would be required. A NetNumber's IETF Internet Draft suggests that ENUM number assignees should be bound by terms and conditions of Tier 1 service providers, including dispute resolution.[125] Like ICANN, this would be top down regulation through contract.

---

[122] This paper does not analyze ICANN's involvement in ENUM. However, it is noteworthy that ICANN has attempted to exercise jurisdiction over ENUM. ICANN, in a recent contract with VeriSign, attempted to exercise control over VeriSign's ENUM activities. ICANN | Information on Proposed VeriSign Agreement Revisions, FAQ # 19 (modified April 1, 2001), *at* http://www.icann.org/melbourne/info-verisign-revisions.htm.

[123] *See also* Shaw, DTI ENUM Workshop, *supra* note 11, slide 14, *at* http://www.itu.int/infocom/enum/dtijune501/dti-june-5-2001-1.PPT; Shaw, ICANN, *supra* note 11, slide 13 (recommending that ENUM infrastructure be "country-neutral" and that transparency is needed "as to clear legal and policy framework, roles, responsibilities, and relationships."), *at* http://www.itu.int/infocom/enum/GACjune1201/gac-june-2-2001-1.PPT.

### Hijacking, Cybersquatting, and Data Authentication

There are several identified naming and fraud problems. These include hijacking, cybersquatting, eavesdropping, and denial of service attacks.[126]

Hijacking or redirection of communication: ENUM numbers query the DNS database for contact information. If access to the NAPTR records is compromised, a third party could alter the contact information. This could result in redirection of traffic away from the desired end point.[127] An example of this would be an ENUM number for a popular call center for the ACME company. The BETA company fraudulently causes the ENUM record to be revised, changing the SIP addresses from ACME to BETA. Now communications go to the BETA call center and BETA attempts to steal ACME's customers.

Eavesdropping: Similar to redirection of traffic, eavesdropping permits the traffic to go through to the desired end point, but only after going through a third party.[128] In this way, the third party can monitor all communications using the ENUM number. For example, communications from CHARLIE to ACME would go through BETA first.

Denial of service: If a company becomes dependent upon traffic directed to it through its ENUM number, and if the security of the ENUM record is compromised, a third party could alter the ENUM record data and effectively block all traffic to the company. This could essentially result in a denial of service attack.

---

[124] Contribution of NeuStar, Inc., US Study Group A Ad-Hoc, *supra* note 4, p. 14.

[125] Ranalli, Tier-1 ENUM, *surpa* note 33, Secs. 4.4, 6.2, *at* http://www.ietf.org/internet-drafts/draft-ranalli-peek-walter-enum-t1roles-01.txt

[126] Ad Hoc ENUM Report, *supra* note 2, Sec. 7.1, *at* http://www.cybertelecom.org/library/enumreport.htm; Ranalli, Tier-1 ENUM, *supra* note 33, Sec. 7, *at* http://www.ietf.org/internet-drafts/draft-ranalli-peek-walter-enum-t1roles-01.txt.

[127] *See* Brown, ENUM Service Reference Model, *supra* note 4, Sec. 8, *at* http://www.ietf.org/internet-drafts/draft-ietf-enum-operation-02.txt. Records could be altered either intentionally and fraudulently or unintentionally or negligently.

[128] Pfautz, ENUM Administrative Process, *supra* note 4, Sec. 5.2, *at* http://www.ietf.org/drafts/draft-pfautz-yu-enum-adm-01.txt

A number of these problems, although not necessarily all, are covered by existing law. For example, if someone hijacked ENUM records, the individual could be in violation of *The Identity Theft and Assumption Deterrence Act*.[129]

In order to respond to these concerns, ENUM services will need to authenticate users and the data submitted. The IETF ENUM convention, again, is that the assignee of a telephone number should be the assignee of an ENUM number. This means that a user's telephone information would need to be authenticated. This could be achieved in a number of ways.

- Directory assistance information for telephone numbers.[130]
- Open Network Architecture, under Computer III, where the Bell Operating Companies are arguably under an obligation to provide this information to enhanced service providers.[131]
- Line Information Database (LIDB).[132]
- Automatic Number Identification where the signaling in the network itself will confirm the callers identity.[133]
- The phone number itself can be called.
- The registrant could be required to show a phone bill.[134]
- Independent authentication or verification through commercial verification services.[135]

---

[129] FTC, *ID Theft: When Bad Things Happen to Your Good Name* (August 2000). *See also* Identity Theft and Assumption Deterrence Act of 1998, Public Law 105-318, 112 STAT. 3007 (Oct 30, 1998); USDOJ, Identity Theft and Identity Fraud (last modified 6/5/2000), *at* http://www.usdoj.gov/criminal/fraud/idtheft.html. Such actions could also be construed as Computer Fraud, 18 U.S.C. §§ 1030.

[130] *See* 47 USC § 222 (giving directory assistance providers rights to list information held by telephone carriers); In re *Provision of Directory Listing Information under the Telecommunications Act of 1934 (sic), as amended*, CC Docket 99-273, First Report and Order, 2001 WL 69358 (Jan. 23, 2001) (clarifying that online databases are directory assistance providers under Sec. 222 and have rights to list information).

[131] *See* Robert Cannon, *Where Internet Service Providers and Telephone Companies Compete: A Guide to the Computer Inquiries, Enhanced Service Providers and Information Service Providers*, 9 COMM. CONSPECTUS 49 (2001).

[132] *See* Kevin McCandless, Illuminent, Number to Name Authentication, SGA Ad Hoc Meeting (March 28, 2001) (advocating LIDB services of Illuminent as solution to data authentication).

[133] *See* Ad Hoc ENUM Report, *supra* note 2, Sec. 7.1, *at* http://www.cybertelecom.org/library/enumreport.htm; Tony Rutkowski, Bryan Mordecai, Approaches to ENUM Implementation in the USA, Dept of State ITAC-T Advisory Committee, SG-A AdHoc Meeting in ENUM, slide 13 (Feb. 12, 2001) (hereinafter Rutkowski, SGA).

[134] Pfautz, ENUM Administrative Process, *supra* note 4, Sec. 5.1, *at* http://www.ietf.org/drafts/draft-pfautz-yu-enum-adm-01.txt

[135] *See* Ad Hoc ENUM Report, *supra* note 2, Sec. 7.1, *at* http://www.cybertelecom.org/library/enumreport.htm; Rutkowski, SGA, *supra* note 133, slide 13 (noting possible use of digital certificate like services); Pfautz, ENUM Administrative Process, *supra* note 4, Sec. 5.1, *at* http://www.ietf.org/drafts/draft-pfautz-yu-enum-adm-01.txt.

There is no indication that currently existing means of authenticating telephone number information is insufficient. In other words, there is no indication that new regulations facilitating assignment are necessary.

### *Telephone Number Issues*

ENUM is a DNS innovation. ENUM numbers are not telephone numbers even though they have the same numerical string as telephone numbers. ENUM presents no telephone number administration issue and will not change the numbering plan.[136]

#### **Numbering Assignment**

ENUM does not affect telephone number assignment. Assignment of public telephone numbers is conducted through the appropriate public telephone authorities. Nothing about ENUM changes this. For all practical purposes, the public telephone network authority does not even have to know that ENUM exists.

Telephone numbers are assigned to telephone network devices so that people can reach them on the telephone network. Assignment of a telephone number for use off of the telephone network makes no sense. If the numerical string is not used on the telephone network, then it is no longer a telephone number. One could no more meaningfully assign a telephone number solely for ENUM purposes than one could assign a telephone number to identify an elephant.

By convention, ENUM numbers are to be assigned according to correlating telephone number assignment. Only assigned telephone numbers would be eligible for ENUM registration. Unassigned telephone numbers would not be assigned.[137] However, if ENUM numbers were assigned that correlate to unassigned telephone numbers, nothing about the assignment would bind the NANP. The assignment of the ENUM number 5551212 to ACME does not give ACME rights to that numerical sting in other contexts; it does not give ACME rights to 5551212 as a telephone number. If the telephone authority assigned 5551212 to BETA, ACME would have no legal rights to challenge this assignment. This is, in effect, the flip side of the argument that ENUM numbers are not telephone numbers. Not only do telephone number regulations not apply to ENUM, but ENUM number assignments do not apply to and do not bind telephone number assignment.

---

[136] *See* NeuStar FAQ, *supra* note 4, p. 1 ("ENUM does not change the Numbering Plan and does not change telephony numbering or its administration in any way. ENUM will not drain already scarce numbering resources because it uses existing numbers.") *at* http://www.enum.org/information/files/enum_faq.pdf; *Id.,* p. 4 ("ENUM will not change the existing right-to-use rules and principles for telephone numbers. ENUM is not intended to change how telephone numbers are administered, but instead facilitate a wide range of applications using phone numbers as subscriber names. ENUM also will not interfere with existing PSTN functions and technology, such as circuit switching, SS7 (ISUP or TCAP), or Intelligent Networking, where similar resource discovery activities are performed through the PSTN legacy technologies."); Shockey, SGA, *supra* note 4, slide 15 (stating "ENUM does not change the Numbering Plan"), *at* http://www.itu.int/infocom/enum/workshopusafeb12-13/shockey.htm.

[137] Contribution of NeuStar, Inc., US Study Group A Ad-Hoc, *supra* note 4, p. 13.

As noted, ENUM numbers and telephone numbers are operationally distinct. If an ENUM number is assigned that correlates to an unassigned telephone number, the ENUM number will still work. The ENUM records would have whatever contact information belongs to the registrant. The fact that the registrant does not have the correlating telephone number does not affect this. Furthermore, as the ENUM query is done entirely over the Internet and not in the telephone signaling network, it would not affect the telephone network.

### Slamming and Cramming

Fraudulent alternations of ENUM records are a concern. However, slamming and cramming, as defined by the FCC, are not. Slamming is the changing of a user's service provider without authorization (i.e., change of long distance service). Cramming is the adding of services without authorization. Neither involves altering the telephone number (the address information) of the user. A person can be slammed (change long distance from AT&T to MCI) and crammed (adding service of call waiting) and no information in ENUM will be changed. Conversely, all of the information in ENUM can be changed without slamming or cramming. ENUM records contain addresses and not information about the services provided for those addresses. The related issues are hijacking, cybersquatting, and DOS attacks, discussed above.[138]

### Number Portability

The IETF has stated that ENUM does not create number portability nor does it create a number portability problem.[139] The assignment of an ENUM number is based on assignment of a telephone number. ENUM therefore needs to authenticate the assignee of a telephone number. Some ENUM supporters assume that authentication will be done by the LEC that serves the telephone customer.[140] If the customer ports the number to another LEC, the source for authentication changes. When a number is ported from Carrier A to Carrier B,

---

[138] *See* discussion on page 29.

[139] Richard Shockey, IETF ENUM Working Group, FAQs About ENUM (Jul. 26, 2000), *at* http://www.ngi.org/enum/pub/DRAFT-SHOCKEY-enum-faq-01.TXT. The statement that ENUM does not affect numbering portability has been noticeably absent from subsequent presentations. *See* Shockey, ITU-T, *supra* note 4, slide 13. NSI / VeriSign also does not view number portability as a crucial ENUM issue. *See* ENUMWorld FAQs (n.d.) *at* http://www.enumworld.com/faqs.html#9. *See also* NeuStar FAQ, *supra* note 4, p. 6 (stating "ENUM is not intended to service this function…"), *at* http://www.enum.org/information/files/enum_faq.pdf.

[140] *See* NeuStar FAQ, *supra* note 4, p. 6 (stating "It is likely that the service provider that allocated the number(s) to the user will be involved in the process of authentication."), *at* http://www.enum.org/information/files/enum_faq.pdf.

Carrier B becomes the holder of the customer information and can verify assignment.[141] AT&T argues that this makes number portability an ENUM issues.

This is incorrect. First, the AT&T scenario describes how number portability affects ENUM, not how ENUM affects number portability. The act of porting a number would change the information source for ENUM, but nothing about telephone number portability has changed.

In addition, as noted above,[142] there are multiple means of verifying number information. The assumption that the LEC serving the customer will be the ENUM source of authentication is not necessarily true.

The ITU is studying the implications of ENUM for number portability; it is believed that the ITU's work will not impact the IETF's ENUM work.[143]

### Non-E164 Numbers (i.e., 911, 711, 411)

How will ENUM handle non-E164 numbers, such as a 911 call? By design, non E.164 numbers would be handled by the device prior to calling the ENUM protocol. If, for example, 911 is dialed, the CPE would set up the call without dipping into the ENUM database.[144] A modern phone is a collection of multiple protocols and programs; not every program is used with each use of the phone. In the case of a 911 call, the ENUM protocol would never be used.

### NANP Number Shortage & New Area Codes

ENUM has no direct impact on the numbering resource;[145] numbering resources are not assigned to ENUM service providers. However, there could be some anticipated indirect impacts.

---

[141] P. Pfautz, IETF Informational Draft, Administrative Requirements for Deployment of ENUM in North America (Sept. 2000), *at* http://www.ietf.org/internet-drafts/draft-pfautz-na-enum-01.txt.

[142] *See* footnotes 130 - 135 and accompanying text.

[143] *See* Liaison, *supra* note 19, *at* http://www.itu.int/infocom/enum/wp1-39_rev1.htm. *See also* Lind, Callflows, *supra* note 4, Sec. 5.3 (noting further work on issue before the ITU), *at* http://www.ietf.org/internet-drafts/draft-lind-enum-callflows-01.txt.

[144] *See* NeuStar FAQ, *supra* note 4, p. 5 (stating "Emergency numbers are generally considered "access codes" and are outside of E.164 and ENUM services. If the user dials an emergency number from a SIP phone, the phone will recognize that it cannot make a SIP connection and will open a gateway to the PSTN."), *at* http://www.enum.org/information/files/enum_faq.pdf. *See also* Contribution of NeuStar, Inc., US Study Group A Ad-Hoc, *supra* note 4, p. 13 (recommending that these types of numbers not be populated into ENUM database).

[145] See Shockey, ITU-T, *supra* note 4, slide 13 (stating "ENUM does not change the Numbering Plan"); Contribution of NeuStar, Inc., US Study Group A Ad-Hoc, *supra* note 4, p. 13-14; *See* NeuStar FAQ, *supra* note 4, p. 1 (stating "ENUM does not change the Numbering Plan and does not change telephony numbering or its administration in any way. ENUM will not drain already scarce numbering resources because it uses existing numbers.").

An indirect pressure could be if ENUM were successful. If ENUM is successful, if many people want ENUM records, and if the one way to have an ENUM record is to have a telephone number, this could create a demand for telephone numbers. Currently a house may have one number but 4 occupants. If each occupant wants an ENUM record, would this mean that the house would now want 4 phone numbers? This could create a drain on the numbering resource.

In addition, ENUM records are frequently referred to as permanent. The assignment of telephone numbers is not. If an individual is known by that individual's ENUM record, that individual may not want to give up the phone number associated with that record. Thus, if the individual sets up a record based on a Virginia phone number, but then moves to California, there is an incentive to keep the subscription Virginia phone number and not recycle it into the numbering pool. This too could create a new demand upon the resource.

### Carrier Selection

ENUM is not about carrier selection.[146] The ENUM database would be populated with address data of various types. Information about the carrier is not included and not relevant. In other words, if the ENUM record reflects that Joe should be reached long distance on a regular telephone at 703-555-1212, it makes no difference in the context of ENUM whether that call is carried by AT&T, MCI or Sprint.

### Telecom Bypass

IETF presentations indicate that ENUM is not about telecom bypass.[147] This is uncertain and indeed contradicted by other IETF presentations.[148] Enabling ENUM seems like an excellent way to provide the originating party options on how to set up the communications; the originating party now has a selection of networks to select from and can now bypass networks the originating party does not desire to use.

### FCC Jurisdiction

As argued above, ENUM numbers are not telephone numbers, they are domain names. The ENUM service is provisioned through the DNS. The policy issues that need to be resolved are issues of DNS administration. Thus, the FCC would not have jurisdiction over ENUM on the grounds that involves telephone numbers.[149]

---

[146] Shockey, ITU-T, *supra* note 4, slide 13; Shockey, SGA, *supra* note 10, slide 15, *at* http://www.itu.int/infocom/enum/workshopusafeb12-13/shockey.htm.

[147] Shockey, ITU-T, *supra* note 4, slide 13.

[148] *See* Lind, ENUM Administration Issues, *supra* note 112, slide 15 (enables "network by-pass").

[149] Section 251(e) gives the FCC authority to "create and designate one or more impartial entities to administer telecommunications numbering and to make such numbers available on an equitable basis." 47 U.S.C. § 251(e). Nothing about ENUM raises an issue related to the administration of telecommunications numbering. Nor would it seem feasible to gain jurisdiction

*Robert Cannon*                                                    ***Draft*** *- Page 34*

*Privacy*

ENUM has the potential to aggregate a tremendous amount of contact information behind a single identifier. This is likely to raise significant concerns.[150] ENUM has been described as an opt-in system.[151] However, there is nothing in the protocol that indicates that ENUM should be an opt-in. Nor is there any known technical reason why it would be limited to an opt in system. Much would depend upon individual business plans. Three business plans can be imagined. ENUM may be implemented at the corporate level so that everyone on a corporate network will have access to contact information and an enhanced ability to contact other people on the network - employees would have *no option* on whether to participate. Second, ENUM may be implemented by a major network as individuals subscribe. One can imagine AOL creating ENUM records as individuals subscribe, utilizing ENUM as a means for members to contact each other. This could be an *opt-out* scenario. Finally, owners of ENUM enabled wireless telephones could, on an individual basis, set up ENUM records. This could be an *opt-in* situation.

There is no limit to the scope of personal information that could be included in the ENUM database. It is conceivable that it could include such things as social security numbers, drivers license numbers, or credit card numbers. No known analysis has been conducted concerning how ENUM complies with the EC Policy on Privacy and Data Protection.[152] Nor has an analysis of privacy implications been provided by privacy public interest organizations or the US Federal Trade Commission.

---

over ENUM under the ancillary jurisdiction doctrine. *See* Statement of Commissioner Michael Powell, Concurring, CC Docket No. 96-98 (n.d.) ("We may eventually win an "ancillary jurisdiction" argument in court against the building owners and landlords, but it does not seem like good policy to propose a new regulatory dictate on these entities before other measures to evaluate the problem or pursue other non-regulatory initiatives prove inadequate."), *at* http://www.fcc.gov/Speeches/Powell/Statements/stmkp917.txt; Separate Statement of Commissioner Michael Powell, Implementation of Sec 255 (Jul. 14, 1999) ("I am unconvinced that such an unrestrained application of ancillary jurisdiction has been sanctioned by the courts, nor do I believe it to be consistent with our own precedents. Accordingly, while I support 99.99 percent of this item and everything that it achieves, I must dissent from its assertion of ancillary jurisdiction.")

[150] Ad Hoc ENUM Report, *supra* note 2, Sec. 7.2 (discussing privacy concerns), *at* http://www.cybertelecom.org/library/enumreport.htm

[151] Ad Hoc ENUM Report, *supra* note 2, Secs. 4.3 & 7.2, *at* http://www.cybertelecom.org/library/enumreport.htm; Faltstrom, ENUM Technical Issues, *supra* note 9, slide 29; Shockey, ITU-T, *supra* note 4, slide 14, *at* http://www.itu.int/infocom/enum/workshopjan01/annex4-shockey.ppt; NeuStar FAQ, *supra* note 4, p. 7 (stating "ENUM would be a subscriber-controlled 'opt-in' system . . . "), *at* http://www.enum.org/information/files/enum_faq.pdf.

[152] NeuStar claims that ENUM is consistent with the EC Privacy Policy. Shockey, ITU-T, *supra* note 4, slide 14, *at* http://www.itu.int/infocom/enum/workshopjan01/annex4-shockey.ppt; Shockey, SGA, slide 16, *at* http://www.itu.int/infocom/enum/workshopusafeb12-13/shockey.htm. However, the conclusion is not substantiated.

## Conclusion

ENUM has the potential to be a tremendous innovation.  Then again, so do many other innovations such as Instant Messaging, SIP, PalmPilots, and the multiple other directory assistance services.  The key policy consideration that ENUM presents is whether it should have government entanglement.  The answer is no.  Not only would it be contrary to pro-competitive policy, not only is there no justification for a government sanctioned monopoly, but government involvement would likely be fatal to the ENUM effort itself, injecting delay and encumbering the project with bureaucracy.  The U.S. Government has long held the policy that it should stay out of the way of the innovation in the highly competitive information technology market; this policy should be maintained.